\documentstyle[11pt,iau185,twoside,epsf]{article}

\begin{document}
\title{Chemical inhomogeneities and pulsation\altaffilmark{1}}
\author{S. Turcotte\altaffilmark{2}}
\affil{Lawrence Livermore National Laboratory, L-413, P.O Box 808,
Livermore, CA 94551, USA}
\altaffiltext{1}{This work was performed under the auspices of the U.S.
Department of Energy, National Nuclear Security Administration by the
University of
California, Lawrence Livermore National Laboratory under contract No.~W-7405-Eng-48.}
\altaffiltext{2}{e-mail: sturcotte@igpp.ucllnl.org}

\begin{abstract}
Major improvements in models of chemically peculiar stars have
been achieved in the past few years. With these new models it has been 
possible to test quantitatively some of the processes involved in the
formation  of abundance anomalies and their effect on stellar
structure. 
The models of metallic A (Am) stars have shown that a much deeper mixing
has to be present to account for observed abundance anomalies. This has
implications on their variability, which these models also reproduce
qualitatively.
These models also have implications for other chemically inhomogeneous
stars such as HgMn B stars which are not known to be variable and
$\lambda$~Bo\"otis stars which can be.
The study of the variability of chemically inhomogeneous stars can
provide unique information on the dynamic processes occurring in many
types of stars in addition to modeling of the evolution of their
surface composition.
\end{abstract}

 
\section{Introduction}

Chemical composition plays an important role in determining the
structure of stars. Its influence comes mostly through the sensitivity
of opacities on the atomic spectra and absorption features of the
elements making up a star.
As most variable stars are unstable
precisely because of how opacity behaves in relation to 
perturbations (the $\kappa$-mechanism), it stands to reason that chemical
composition be of fundamental interest when studying pulsating stars.

Not only does the effect of the bulk metallicity of stars need
to be considered
but the effect of composition gradients ({\sl a.k.a.}~chemical
inhomogeneities) on stellar pulsations has to be investigated, and that
has been done only crudely so far. The difficulty lies in that in order to model 
chemically inhomogeneous stars properly it is
indispensable to be able to calculate opacities correctly.
Standard opacity tables can follow gradients of the major constituents
but not of individual elements. While valid in most cases these tables
are not adequate in stars in which substantial departures from a
homogeneous composition occur and in which the total metallicity $Z$
does not reflect accurately the opacity of the actual chemical
composition everywhere in the star.

In recent years some models have tackled those issues.
Charpinet et al.~(1996), for example, have used an approach
in which the influence of the variation of a specific element, iron in
their case, is followed.
They were able to predict that sdB stars should be variable and that
diffusion of iron was able to provide the necessary abundance and opacity
enhancements.
Another more ambitious approach is to model stars in a
consistent fashion using detailed monochromatic opacities for all
important elements and a full
treatment of diffusion (Richer et al.~2000 and references therein).
Such models allow a detailed study of chemically
inhomogeneous stars both in terms of surface abundance anomalies and
in terms of stellar pulsations for a wide range of stellar types and 
evolutionary stages.

Simultaneously, the need for such work has been underlined by the 
increasing number and types of chemically peculiar stars (hereafter CP
stars) which are now recognized as variable.
In the past, there was an apparent dichotomy between pulsations and
peculiar compositions in stars with only a few exceptions (Kurtz~1978).
 Models of the time, including only the
effect of hydrogen and helium abundance gradients, were mostly in 
agreement with the observations (Cox et al.~1979).
The discovery of new classes of variable CP stars, roAp stars by Kurtz~\&
Wegner~(1979) and $\lambda$~Bo\"otis stars by Gonzalez et al.~(1998),
and the constant additions to the $\delta$~Delphini class of variable evolved
metallic A stars (Kurtz~2000 and references therein) showed the dichotomy to
be at least partly an illusion.

\section{Chemical inhomogeneities in stars}

CP stars are identified and defined by their observed surface
abundances. However the connection between surface abundances and
internal abundance profiles depends on how and where mixing occurs in
the star as well as which process or processes are causing the
surface chemical peculiarities. As stellar pulsations depend on the
internal profile, one must be wary of jumping to conclusions based
solely on surface abundances when studying CP stars.

The most prevalent ways of modifying the surface composition in a star
are: 1) differential effects of microscopic diffusion (Richer et
al.~2000), 2) dredge-up of matter from the deep interior through mixing 
or mass loss (Fowler et al. 1965), 3) accretion of
matter affected by dust formation (Venn~\& Lambert~1990) 
or nucleosynthesis (Guthrie 1971), 
4) differential mass loss where some elements are ejected and others not
(Babel~1996).
The connection between the surface abundances and internal profiles
are essentially determined by whatever mixing exists in the superficial
regions of the star for all these processes. Some of these processes
however are expected to be active mostly in young stars (accretion)
while others are expected to have an effect only later on (dredge-up).

It is assumed that convective motions are fast enough to 
mix convection zones instantaneously, thereby homogenizing its
composition and diluting any influx or out flux of matter in its
entirety. In addition there is ample evidence that further mixing occurs
in the stable region below the convection zone. The composition observed
at the surface extends therefore to the base of this fully mixed region.
In cool stars this region is very deep, in hot stars there might not be
any significant mixed region at the surface. Below the mixed region the
abundances can be very different than those observed at the surface.
As a result, surface abundances are a test of the depth of the
superficial mixing while seismology can potentially allow the study of
mixing below this mixed region. 

\section{Pulsations in CP stars}

As the chemical composition changes in a star so will the opacity
profile. Variations of the composition will affect mainly the $\kappa$-mechanism
for driving pulsations although they may have indirect consequences for
other mechanisms, such as the one proposed for $\gamma$~Doradus stars
(Guzik et al.~2000). I will concentrate here on stars were the $\kappa$-mechanism
dominates.

There are three regions in a star where opacity gradients occur which
can lead to the excitation of pulsations. There is one due to the
ionization of HI at low temperatures (roughly 10\,000~K), one
due to the ionization of HeII (40\,000~K) and the last one
due to the opacity of heavy elements, mainly iron-peak elements, 
at around 2000\,000~K. Not all three contribute significantly to
excitation or damping in all stars but in CP stars one can expect
large variations in their relative contributions to the net
excitation of pulsations compared to chemically normal stars. 

\subsection{Diffusion and pulsations}

Diffusion has been shown to have a major effect on the composition of slowly
rotating A and the cooler B stars. Of the types of stars for which diffusion has
the largest effects two are particularly interesting in the present
context. Am stars are A type stars enriched in iron-peak elements 
and HgMn stars are their more massive (B type stars) counterparts which
feature varying composition anomalies especially in manganese and/or mercury.
There is a strong dichotomy between pulsations and the CP phenomenon in
those stars, except for the aforementioned $\delta$~Delphinis  
which are thought to be evolved Am stars. 

The A stars have been modeled by Richer et al.~(2000) and their
stability was investigated by Turcotte et al.~(2000). 
The major effect of diffusion in A stars in terms of pulsations is to reduce the helium
abundance in superficial regions, thereby decreasing the excitation of
the modes typical of variable A stars. The models however show that superficial 
mixing must be very deep for the models to
reproduce surface abundance anomalies. The consequence of this deep
mixing is to preserve some helium in the driving region. 

\begin{figure}
\begin{center}
\mbox{\epsfxsize=0.7\textwidth\epsfysize=0.8\textwidth\epsfbox{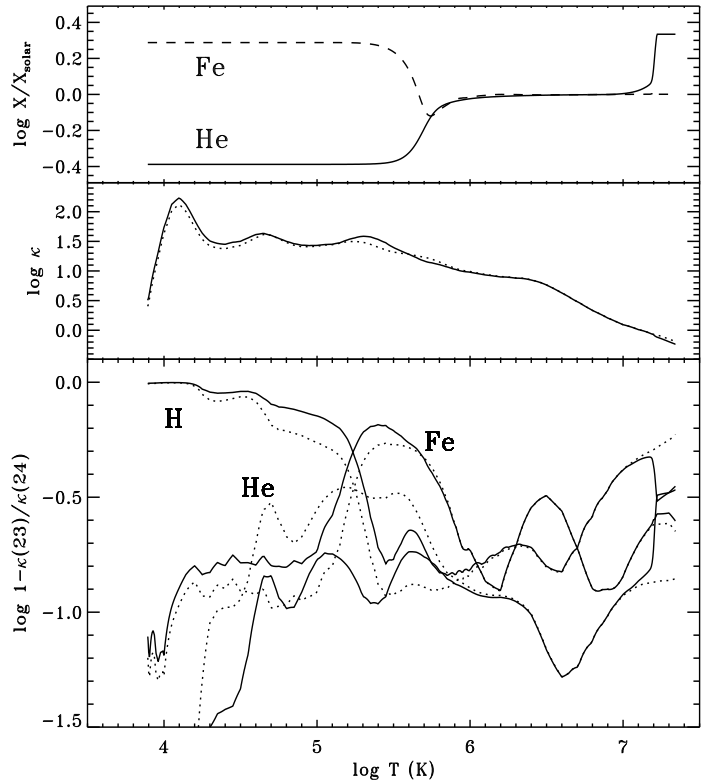}}
\caption{
A model with diffusion (solid line) and one without (dotted) are
compared. The top panel shows the abundance profiles of He and Fe as a
result of diffusion relative to the homogeneous model. 
The middle panel shows the Rosseland opacity for both models. 
The bottom panel shows the contribution of H, He and Fe to the total
opacity ($\kappa(24)$; $\kappa(23)$ is the opacity with one
element removed).
%This figure illustrates the effect of diffusion on the
%opacity profile in the model of a 1.80~M$_\odot$ star with 
%a surface composition representative of Am stars. The top panel
%shows the relative abundance profiles of helium and iron to the solar abundance, 
%the middle panel
%shows the opacity profile in the model with diffusion (solid line) and a
%homogeneous reference model (dotted line), and the last panel shows the
%contribution of the elements identified to the opacity in both
%models (solid line: Am star, dotted line: reference model). $\kappa(23)$ represents
%the opacity calculated with all the elements {\sl except} the element
%identified, $\kappa(24)$ is the opacity calculated with {\sl all} elements included.
\label{fig:opacity}}
\end{center}
\end{figure}

Figure~\ref{fig:opacity} shows the relative contribution of major
elements in a star of 1.8~M$_\odot$ as a function of temperature.
Two models are plotted, one with a standard solar composition
(Grevesse \& Noels~1993), and one representative of an Am star.
It shows that the opacity in the driving region for
$\delta$~Scuti-like pulsations (HeII) actually increases as the result
of diffusion but that the contribution of helium decreases. As the
driving is provided by the opacity ``hump'' due to helium at 40\,000~K,
the relative disappearance of helium levels the opacity gradient and
removes much of the driving. The more helium settles out of this region,
the less likely is the star to become variable. The increase in the
contributions of hydrogen and metals play little role in these stars.
However, there are suggestions that metals may contribute significantly to the 
variability of Ap stars (Michaud et al., these proceedings).

\begin{figure}
\begin{center}
\mbox{\epsfxsize=0.6\textwidth\epsfysize=0.6\textwidth\epsfbox{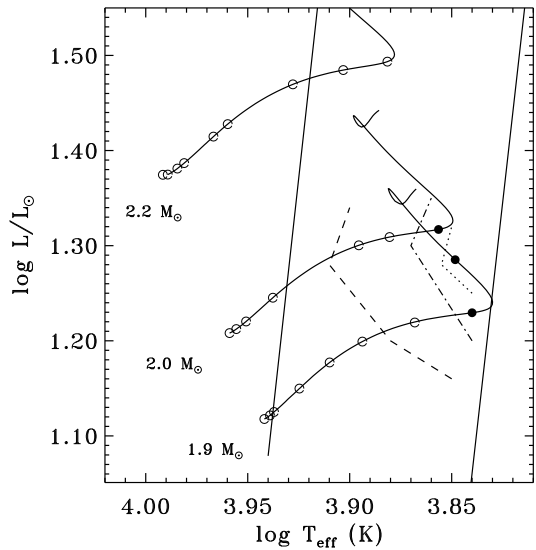}}
\caption{The approximate edges of the instability strip are shown for
$\delta$~Scuti stars (solid) and Am stars (dash-dotted). The blue edge
for models with deeper (dashed) and shallower (dotted) turbulence are
also shown. Filled circles indicate models of variable Am stars, open
circles indicate stable Am stars. The evolutionary path for models of Am
stars of 1.9, 2.0 and 2.2~M$_\odot$ are shown.
%The instability strip for $\delta$~Scuti stars is shown in
%the HR diagram by the two solid lines. An estimate of the blue edge of the
%instability strip for Am stars is shown by the dash-dotted curve. The
%dashed curve represents the limit of variability for models with deeper 
%turbulent mixing while the dotted curve shows the blue edge for models 
%with shallower mixing. The open circles show the models of Am stars for which all modes are stable, the
%filled circles show the models in which at least one mode was found
%to be overstable. The evolutionary path for models of Am stars of 1.9, 2.0 and
%2.2~M$_\odot$ are shown.
\label{fig:IS}}
\end{center}
\end{figure}

The stability analysis of the models showed that pulsating Am stars are restricted to
fairly evolved stars, in agreement with observations, and that the blue edge 
of their ``instability strip'' is sensitive to the depth of the mixing
(Figure~\ref{fig:IS}). As the blue edge shifts toward the red edge the
less helium remains in the driving as a result of mixing, the region of
instability also shifts to more massive stars. The limits are far from
being well defined and the models do not test whether there is an upper
mass limit to the instability region.

If diffusion in A stars tends to stabilize stars to pulsations and
observations and theoretical predictions are overall in accord,
the situation in the B stars is more confused. Models have shown that 
iron-peak elements accumulate both at the surface and in a 
localized region at a temperature of roughly 200\,000~K where they
dominate the opacity. All the models show a significant increase in
the opacity in that region, which can lead to the emergence of localized
convection in some cases.
Not coincidentally, pulsations in B stars are driven by the opacity of metals in that
region. Diffusion should therefore lead to an increase in
driving. HgMn stars could be expected to be variable to the same degree 
as Slowly Pulsating B (SPB) stars which share the same region of the H.-R. diagram but the
opposite is observed. No HgMn stars has been found to be variable
because of pulsations.

In order to prevent pulsations in HgMn stars one would need to either
introduce damping in the superficial regions or reduce the opacity of
metals in the driving region. Introducing mixing would not solve the
problem since normal metal abundances lead to ``normal'' SPB stars.
No significant damping due to abundance variations of other elements,
notably helium, were found in preliminary models. It remains that the
HgMn phenomenon seems well correlated with slow rotation which may well
give a hint of the solution of this conflict between the and
observations. 

\subsection{Accretion and pulsations}

Matter accreted at the surface of a star will be redistributed 
within the superficial mixed region and may diffuse below it. 
If the mass accreted is large enough the entire mixed zone will
take on the chemical composition of the accreted matter
(Turcotte~\& Charbonneau~1993).
 
Accretion is the leading mechanism to explain the observed composition
of $\lambda$~Bootis stars (Solano et al.~2001).
If accretion indeed is at the root of the $\lambda$~Bootis phenomenon 
then one expects these stars
to be young, with ongoing accretion (Charbonneau~1993, Turcotte~\&
Charbonneau~1993).
Although some observations play in favor of the accretion
scenario, such as evidence of circumstellar matter and
even infall in some cases as well as their $\delta$~Scuti-type
variability, there remain some problems.
The most troublesome problem is that the $\lambda$~Bootis stars
cover can be ZAMS or TAMS stars or anywhere in between. 

The Richer et al.~(2000) models for A stars have shown that deep mixing
can be expected in slowly rotating stars. That mixing is expected to be
at least partially driven by rotation. It follows that one may expect
that faster rotating stars such as $\lambda$~Bootis stars have 
mixing below the convection zone extending as deep or even deeper than
in Am stars. 
Deep mixing in $\lambda$~Bo\"otis stars has a few repercussions on the 
mechanics of accretion and on the modeling of the stars.
\begin{figure}
\begin{center}
\mbox{\epsfxsize=0.6\textwidth\epsfysize=0.6\textwidth\epsfbox{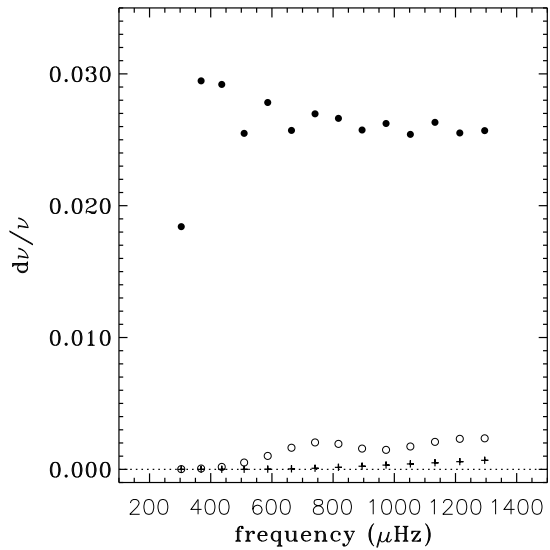}}
\caption{Predicted frequency differences of radial modes in models of a
1.80~M$_\odot$ star with $\lambda$~Bo\"otis-like superficial abundances
relative to a normal star.
The crosses are for a model where the chemical anomalies attain a depth of
$\log T = 4.4$ (just below the SCZ in standard models of A
stars), the open circles for $\log T=5.0$ (1000 times more massive),
and the filled circles for $\log T=5.8$ ($10^6$ times more massive).
%Predicted frequency differences of radial modes in models of a
%1.80~M$_\odot$ star with $\lambda$~Bo\"otis-like superficial abundances
%relative to a chemically. The abundances are one tenth of the solar
%abundances by mass for all elements except: He,C,N,O,S which are solar and
%H which is increased to compensate for the missing metals.
%The models differ by the depth to which the
%abundances of heavy elements are anomalous. The
%crosses are for a model where the chemical anomalies attain a depth of
%$\log T = 4.4$ (just below the bottom of the convective zone in standard models of A
%stars), the open circles for $\log T=5.0$ (1000 times more massive), 
%and the filled circles for $\log T=5.8$ ($10^6$ times more massive). 
%Modes of higher degree follow the same trends.
\label{fig:lboo}}
\end{center}
\end{figure}

As accreted material must be diluted in the entire mixed
region, then the mass which needs to be accreted for surface abundances
to reflect those of the accreted matter must be proportionally larger.
It also means that the time scale for the establishment of the characteristic
surface composition increase as well which might imply that accretion rates are higher
than previously thought. The peculiar surface composition
will persist longer after accretion is halted than the few Myr obtained by Turcotte \&
Charbonneau~(1993). Finally, and perhaps more importantly in the context
of pulsations, the star will have a peculiar composition to a much
greater depth than in anterior models. Such large abundance anomalies
to such large depths can have non negligible effects on stellar
structure.

If one recalls the models of Am stars, the chemically homogeneous region
extends from the surface to points beyond the region where iron-peak
elements dominate the opacity (at roughly 200\,000~K). As
$\lambda$~Bo\"otis stars are superficially metal-poor, the opacity in
the metal bump will be significantly lowered. Although this has some
effect on driving, the driving is still dominated by helium as in Am
stars. The structure of the star is changed somewhat and the frequencies
of the modes of oscillation in the star are changed by a significant
amount as shown in Figure~\ref{fig:lboo}. 
The effect is not significant if mixing is shallower.

If mixing is indeed as deep as the models of
Am stars suggest, then models including opacities adequate for 
the peculiar chemical composition may be necessary if one hopes
to constrains the models or analyze the seismology of these stars in
more details. 

\section{What can we learn from variable CP stars?}

The evolution of the chemical composition in a star is very sensitive to
mixing processes occurring within it. As there is an obvious signature
of separation processes likely modified by mixing in chemically peculiar
stars, it follows that they are objects through which one stands of
learning a great deal about those mixing processes. In so-called normal
stars, one does not have the same tools to study those processes.

We have shown here that detailed modeling of A stars has lead to a
radical change in the paradigm for their structure, pointing to
much deeper mixing than was expected from anterior models. 
These models also reproduce qualitatively the observed variability in
metallic A stars. Detailed models of specific stars will be able to
confirm whether the predicted pulsations agree with observations as well
as predicted surface composition does to what can be measured in Am
stars.

In B stars, there is an apparent contradiction between theoretical
expectations for pulsation driving and observed variability in
HgMn stars. Solving this problem is likely to bring about a 
better understanding of mass loss and mixing in B stars.
Better models of CP B stars will be possible when new opacity data
better suited to calculate radiative pressure on individual elements
in superficial regions are released by the OPAL group.

As models of CP stars improve, models of other variable stars will be
achieved such as roAp stars for which preliminary models are
now being attempted (see Richard et al., and Michaud et al. in these
proceedings).

\section*{Discussion}
\parindent=0pt
 
{\it C. Aerts~:} Some of SPBs studied by De Cat (2001) have deviating
chemical abundance (mostly in Si). They all have low $\sin i$.
How does that compare to your excitation models ?\medskip

{\it S. Turcotte~:} Surface abundance anomalies in B stars are very
difficult to interpret in terms of internal abundance profiles because the structure
and mixing or separation processes occurring in the external regions
are essentially unknown. 
However low $\sin i$ in SPB stars is an intriguing problem if the
stars are really slow rotators because one would then have to introduce another
parameter in addition to rotation to understand the non variability of HgMn stars. 

\end{document}